\documentclass[aps,prb,twocolumn,showpacs,floatfix,groupedaddress,superscriptaddress]{revtex4-2}
\pdfoutput=1
\usepackage[linkcolor=blue,colorlinks=true,breaklinks=true,citecolor=blue,urlcolor=blue]{hyperref}
\usepackage{graphicx,graphics}  
\usepackage{dcolumn}   
\usepackage{braket}
\usepackage[mathscr]{euscript}
\usepackage{amsbsy}
\usepackage{bm} 
\usepackage{amssymb} 
\usepackage[T1]{fontenc}
\usepackage{amsmath}
\usepackage{xcolor}
\usepackage[caption=false,labelformat=simple]{subfig}

\makeatletter
\newcommand{\phantomlabelabovecaption}[2]{
	\protected@write\@auxout{}{
		\string\newlabel{#2}{
			{\number\numexpr\thefigure+1\relax#1}{\thepage}
			{\number\numexpr\thefigure+1\relax#1}{#2}{}
		}
	}
	\hypertarget{#2}{}
}
\makeatother

\begin{document}

\title{Magneto transport of pressure induced flatbands in large angle twisted bilayer graphene}
\author{Ayan Mondal}
\email{ayanmondal367@gmail.com}
\affiliation{Department of Physical Sciences, Indian Institute of Science Education and Research Kolkata\\ Mohanpur-741246, West Bengal, India}
\author{Priyanka Sinha}
\thanks{{\bf Present Address:} Institute of Physics of the Czech Academy of Sciences (FZU), Na Slovance 1999/2, 182 00 Prague 8, Czechia}
\email{sinhapriyanka2012@gmail.com}
\affiliation{Department of Physical Sciences, Indian Institute of Science Education and Research Kolkata\\ Mohanpur-741246, West Bengal, India}
\author{Bheema Lingam Chittari}
\email{bheemalingam@iiserkol.ac.in}
\affiliation{Department of Physical Sciences, Indian Institute of Science Education and Research Kolkata\\ Mohanpur-741246, West Bengal, India}

\date{\today}

\begin{abstract}
Twisted bilayer graphene (TBG) exhibits flat electronic bands at the so-called magic angle ($\sim 1.1^\circ$), leading to strong electron correlations and emergent quantum phases such as superconductivity and correlated insulating states. However, beyond the magic angle, the band structure generally remains dispersive, diminishing interaction-driven phenomena. In this work, we explore the equivalence between pressure-induced flatbands and the magic-angle flatband in large-angle TBG by systematically analyzing the role of interlayer coupling modifications under perpendicular pressure.  We show that pressure-induced flatbands exhibit spatial localization similar to magic-angle TBG, with charge density concentrated in the AA-stacked regions. Furthermore, the Hall conductivity and magneto-transport properties under an external magnetic field reveal that these pressure-induced flatbands share key signatures with the quantum Hall response of magic-angle TBG. The obtained Hofstadter spectrum shows four consistent low-energy gaps across all twist angles under pressure, which align with the calculated Hall conductivity plateaus. Our findings suggest that pressure offers an alternative pathway to engineer flat electronic bands and correlated states in TBG, extending the landscape of tunable moiré materials beyond the constraints of the magic angle.
\end{abstract}
%%
%\pacs{72.80.Vp, 73.20.At, 73.22.Gk,}
\maketitle
%%
%%
%%-----------------------------------INTRODUCTION----------------------------------------------

\section{Introduction}
Twisted bilayer graphene (TBG) \cite{Santos,Bistritzer,Sboychakov,Andrei} has emerged as a fascinating platform for exploring correlated electron physics due to its moiré-induced flat bands and tunable electronic structure. At the magic angle ($\sim 1.1^\circ$), these flat bands enhance electron interactions, leading to strongly correlated phases \cite{Tarnopolsky,Kennes} such as superconductivity \cite{Cao1}, Mott-like insulators \cite{Cao2}, and unconventional quantum Hall effects \cite{Serlin}. However, the electronic properties of TBG are not solely dictated by the twist angle—external parameters such as electric fields \cite{Talkington,Sinha}, strain \cite{Hou,Qiao,Huder,Zhang, Li}, and pressure \cite{Yankowitz1,Yankowitz2,Carr,Lin,Chittari} provide additional control over its band structure. Recently~\cite{Sinha}, we investigated the electronic and magnetotransport properties of TBG under external electric and magnetic fields. We found that at small twist angles, the Hall conductivity transitions from a half-integer quantum Hall effect $\sigma_{xy} = \pm~ 4(n+\frac{1}{2})(2e^2/h)$ to an integer quantum Hall effect $\sigma_{xy} = \pm~ 2n(2e^2/h)$ ($n=0,\pm~ 1,\pm~ 2,…$), around $\theta = 2.005^\circ$, where a well-defined zero-energy Hall plateau appears. However, this plateau exhibits a kink with a nonzero longitudinal conductivity. The application of an electric field makes the flat bands more dispersive, reducing electron localization, while the flat bands remain largely unchanged under a magnetic field. Additionally, the application of an electric field modifies the Hall conductivity plateaus, with low-angle TBG showing a reduction in the width of the zero-energy plateau, whereas at high angles, new plateaus emerge, increasing the number of Landau level peaks within the same energy range. Importantly, we observe that as the twist angle increases, TBG progressively behaves like two decoupled monolayer graphene sheets. These results highlight the intricate interplay between twist angle, external fields, and quantum transport in TBG. Additionaly, previous theoretical \cite{Carr,Lin,Chittari,Romanova,Ge,Padhi,Wang} and experimental \cite{Yankowitz1,Yankowitz2} studies have demonstrated that pressure can induce flat bands in TBG by modifying interlayer coupling, leading to enhanced correlation effects. However, the extent to which these effects persist at significantly higher twist angles, as well as whether key features such as electron localization, quantum Hall plateaus, and moiré-induced transport phenomena remain intact, is not well understood. Large-angle TBG generally exhibits dispersive bands and weak electron correlations, but applying pressure modifies the interlayer coupling, potentially altering its electronic structure in unexpected ways. 
\begin{figure*}[t!]
    \phantomlabelabovecaption{(a)}{fig:model_1}
    \phantomlabelabovecaption{(b)}{fig:model_2}
    \phantomlabelabovecaption{(d)}{fig:model_4}
    \includegraphics[width = 0.9\textwidth]{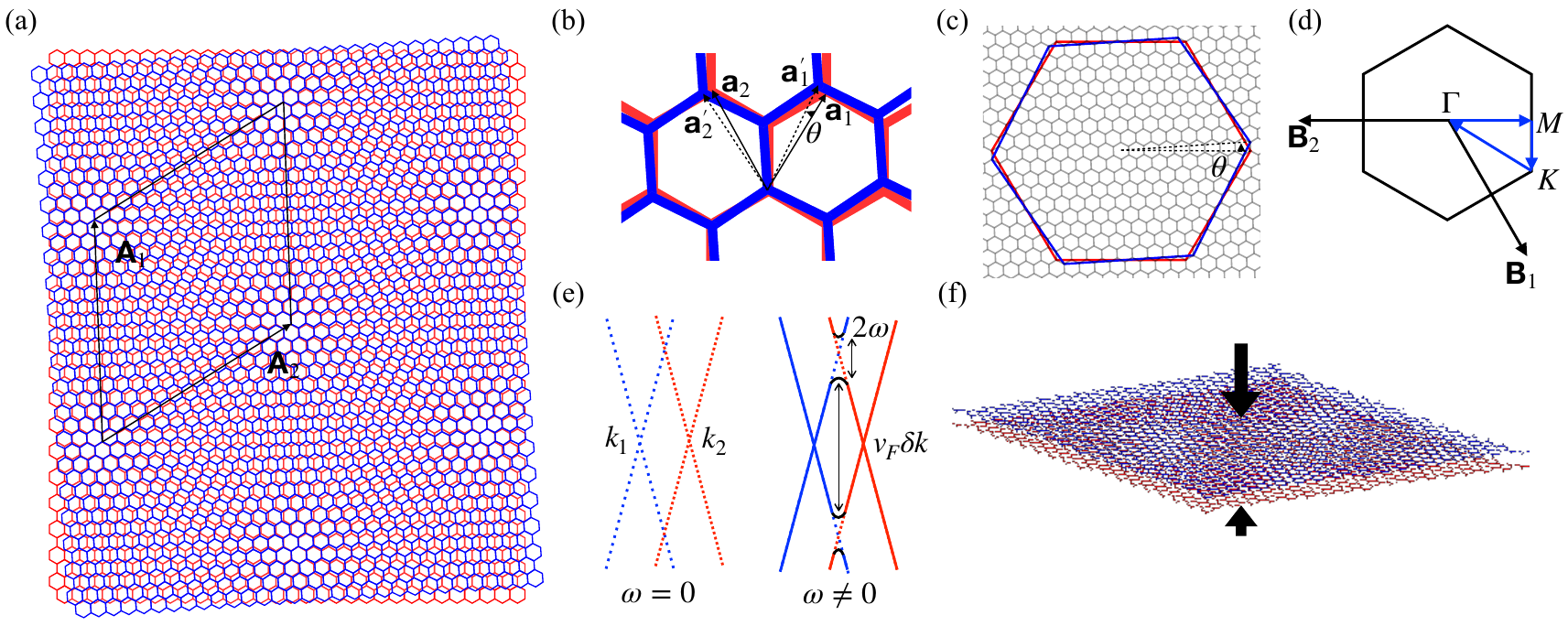}
    \caption{(color online)(a) Schematic of twisted bilayer graphene (TBG) with a commensurate twist angle of $3.890^\circ$. The top layer (blue) is rotated counterclockwise while keeping the bottom layer (red) fixed. The moire unit cell is defined by lattice vectors $\Vec{A}_1$ and $\Vec{A}_2$ (black rhombohedral outline). (b) Unit cell diagram of the bottom and top graphene layers with lattice vectors $\Vec{a}_1,\Vec{a}_2$ (bottom layer) and $\Vec{a}_1^\prime,\Vec{a}_2^\prime$(top layer), showing the relative twist angle $\theta$. (c) The relative twist in reciprocal space results in a smaller moire Brillouin zone (black). (d) Moire Brillouin zone with reciprocal lattice vectors are $\Vec{B}_1$ and $\Vec{B}_2$. The path $K \rightarrow \Gamma \rightarrow M \rightarrow K$ represents the high-symmetry route for band structure calculations. (e) Schematic of two interacting Dirac cones from different layers, where interaction separates high-energy bands from low-energy bands. The bandwidth of the low-energy bands scales with $v_{F}\delta K$, where $v_{F}$ is the Fermi velocity and $\delta K = K_1 -K_2$ is the separation between Dirac points. (f) Representation of TBG under perpendicular pressure, highlighting the system's response to external forces.}
    \label{fig:model}
\end{figure*}
\par In this paper, we aimed to explore the impact of pressure on large-angle TBG, focusing on commensurate higher twist angles. By investigating how pressure influences band flattening, electron localization, and quantum transport properties across different twist angles, we aim to determine the extent to which magic-angle-like behavior can be extended beyond the conventional regime. This study provides a systematic approach to understanding the role of pressure in engineering electronic states in twisted graphene systems and contributes to the broader field of moiré materials and strongly correlated physics. In Sec.~\ref{II}, we provide a detailed discussion of the geometry and model Hamiltonian. Section~\ref{III} presents our results, beginning with Sec.~\ref{A}, where we analyze the band structure of pressure-induced flatbands. In Sec.~\ref{B}, we explore the spatial localization of electrons in these flatbands. The effect of a magnetic field on the flatbands is examined in Sec.~\ref{C}. Section~\ref{D} discusses the low-energy Hall plateaus under perpendicular pressure, while Sec.~\ref{E} establishes the correspondence between these Hall plateaus and the Hofstadter butterfly spectrum. Finally, in Sec.~\ref{IV}, we summarize our findings.

\section{Model} \label{II}
The geometry of twisted bilayer graphene (TBG), as described ib Fig.~\ref{fig:model_1} which illustrates the moiré pattern formed due to the relative rotation of two graphene layers. The top layer (blue) of AA-stacked bilayer graphene is rotated counterclockwise while keeping the bottom layer (red) fixed, with the center of rotation at an A sublattice site. The lattice vectors for the unrotated layer are $\Vec{a}_1=(1/2,\sqrt{3}/2)a$ and $\Vec{a}_2 = (-1/2,\sqrt{3}/2)a$ where $a=2.46  $\AA~is the lattice constant of graphene. After applying a twist angle $\theta$, the lattice vectors of the rotated layer become $\Vec{a}_1^{\prime}$ and $\Vec{a}_2^{\prime}$ as shown in Fig.~\ref{fig:model_2}. For commensurate structures, the twist angle $\theta$ satisfies the relation \cite{Shallcross}:
\begin{equation}
    cos(\theta) = \frac{m^2+n^2+4mn}{2(m^2+n^2+mn)}
\end{equation}
where m,n are integers, For example, m=n=1 gives $\theta=0^\circ$(AA-stacking) and m=1,n=0 gives $\theta = 60^\circ$(AB-stacking). This twist-induced lattice mismatch generates a larger periodic unit cell (supercell) in TBG, with lattice vectors: $\Vec{A}_1 = n\Vec{a}_2+m\Vec{a}_1$ and $\Vec{A}_2 = -m\Vec{a}_2+(n+m)\Vec{a}_1$. A supercell contains $m^2+n^2+mn ( \equiv n_0)$ A and B sublattice sites in each layer, resulting in a total of $4n_0$ sites across both layers. In reciprocal space, the lattice mismatch between the two layers forms a mini Brillouin zone for TBG, as depicted in Fig.~\ref{fig:model_4}. \\

We have considered the minimum tight-binding Hamiltonian for twisted bilayer graphene \cite{Xianqing}. This tight binding model is able to produce the highest magic angle near $\sim 1.084^{\circ}$ and is consistent with experimental data \cite{Cao1, Cao2}. Like all other layered systems, the Hamiltonian consists of two parts - intralayer($H_{\parallel}$) and interlayer($H_{\perp}$):
\begin{align}
    H=&H_{\parallel}+H_{\perp} \nonumber \\
    =&- \sum_{\substack{i \ne j \\ m}}t_{ij}^{mm}(\hat{a}_{m,i}^{\dagger}\hat{a}_{m,j}+H.c.) \nonumber \\
    & - \sum_{\substack{i , j \\ m}}t_{ij}^{m,m+1}(\hat{a}_{m,i}^{\dagger}\hat{a}_{m+1,j}+H.c.)
\end{align}
For twisted bilayer graphene the layer index($m$) can take values 1,2. $t_{ij}^{mn}$ denotes the hopping integral between two sites, $i$ from layer $m$ and $j$ from layer $n$. The operators $\hat{a}_{m, i}^{\dagger}$ and $\hat{a}_{m, i}$ denote creation and annihilation operators of a $p_{z}$ state of carbon atom respectively at site $i$ in layer $m$. H.c. denotes the Hermitian conjugate term.
\begin{figure*}[!ht!]
    \phantomlabelabovecaption{(a)}{fig:bs_1}
    \phantomlabelabovecaption{(b)}{fig:bs_2}
    \includegraphics[width = 0.9\textwidth]{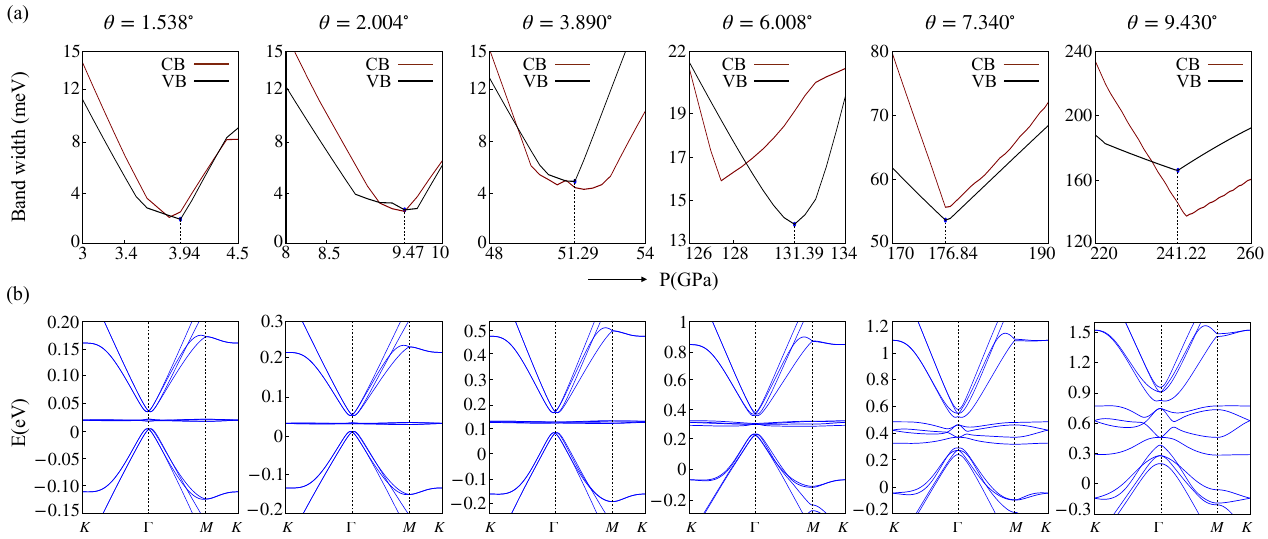}
    \caption{(color online)(a) Variation in the bandwidth of the highest energy valence band (black) and the lowest energy conduction band (brown) upon applying perpendicular pressure for twisted bilayer graphene at the commensurate angles $1.538^\circ$, $2.004^\circ$, $3.890^\circ$, $6.008^\circ$, $7.340^\circ$, and $9.430^\circ$. The pressure at which the valence band reaches its minimum is indicated by a dotted line, with the corresponding pressure values labeled along the x-axis. (b) Band structure (blue) in the superlattice Brillouin zone along the high-symmetry path $K-\Gamma-M-K$ for each of the above twist angles, shown at the specific perpendicular pressure where the minimum in the highest valence band occurs.}
    \label{fig:bs}
\end{figure*}
For intralayer hopping, we have limited to the first nearest neighbor with $t_{<ij>}^{mm} = 3.09 eV$ which reproduces the Fermi velocity $v_{F} \approx 1 \times 10^{6} m/s$ for monolayer graphene \cite{Neto}. Interlayer hopping integral is considerebd to be isotropic in this case and can be written as a function of mutual distance vector between two atoms from different layers projected on a single layer($r$) :
\begin{equation}
    t(r)= V_{pp \sigma}^0 e^{-(\sqrt{r^{2}+d^2}-d)/\delta}\frac{d^{2}}{r^{2}+d^{2}}
\end{equation}
It will be maximum with value $V_{pp\sigma}^{0} = 0.39 eV$ when $r=0$. In that case, one of the atoms from the top layer will be atop another atom from the bottom layer separated by interlayer distance $d$. The hopping strength between two atoms with $r>0$ will decrease exponentialy multiplied by the cosine of tilting angle \cite{Moon}. Here, $\delta$ defines the cutoff for $r$ to neglect interactions at large distances. For $\delta = 0.27 $\AA~ we can reproduce AB bilayer graphene band structures, indicates the inclusion of all the significant contribution to hopping energy. The perpendicular distance between two layers is $d = 3.35 \AA$. We have considered zero onsite energy for each atom.

Following the reference \cite{Chittari}, by assuming an approximately equal distribution of AB, BA and AA stacking areas and averaging the values of interlayer tunneling at the Dirac point for different stacking configurations, a polynomial fit for the correlation between the interlayer coupling parameter $\omega = V_{pp\sigma}^0/3$ and pressure $P$ is established as $P=A\omega^{2}+B\omega+C$ with numerical parameters $A=324.7 GPa(eV)^{-2}$, $B=-35.47 GPa(eV)^{-1}$ and $C = -0.4671 GPa$ for interlayer distance $3.35 $\AA~ for out of plane LDA relaxations \cite{Jung1}.

\section{Results and Discussions}\label{III}
\subsection{Pressure induced flatbands}\label{A}
In this section, we will discuss the effect of low-energy bands in the presence of uniform perpendicular pressure. In Fig.~\ref{fig:bs}, we illustrate that by applying perpendicular pressure, a minimum in the bandwidth can be achieved in both low-energy bands for any commensurate twist angle. We present results for six specific commensurate angles: $1.538^\circ$, $2.004^\circ$, $3.890^\circ$, $6.008^\circ$, $7.340^\circ$, and $9.430^\circ$. While additional commensurate angles exist, defined by ($m,n$), we aim to capture the full range of twist angles by examining these six representative cases. In Fig.~\ref{fig:bs_1}, we plot the bandwidth of the two lowest energy bands—the highest valence band (shown in black) and the lowest conduction band (shown in brown)—as a function of pressure, measured in GPa. Dotted lines indicate the pressure required to minimize the bandwidth of heighest valance band. Throughout this paper, we refer to the pressure at which the minimum bandwidth occurs in the highest valence band as the pressure needed to achieve the flattest band attainable by applying perpendicular pressure. The minimum bandwidth for the highest valence band occurs at different pressures depending on the twist angle: 3.94 GPa for $1.538^\circ$, 9.47 GPa for $2.004^\circ$, 51.29 GPa for $3.890^\circ$, 131.39 GPa for $6.008^\circ$, 176.84 GPa for $7.340^\circ$, and 241.22 GPa for $9.430^\circ$. At the magic angle of $1.084^\circ$ (Appendix~\ref{app:magic_flat}), however, both low-energy bands reach their minima simultaneously. At larger twist angles, this simultaneous minimum no longer occurs, and as the twist angle increases, the differences between the minima in two low energy bands become more pronounced. This indicates that pressure does not affect the two bands in a symmetric way at higher angles. In Fig.~\ref{fig:bs_2}, We have plotted the corresponding band structure along the high-symmetry path $K-\Gamma-M-K$ in moiré momentum space, where the bandwidth minima occur for each of the six commensurate twist angles. The band structures are plotted within specific energy limits to focus on the lowest energy bands, as the higher energy bands almost unaffected by the application of pressure. For low twist angles, extremely flat bands can be achieved by applying perpendicular pressure. For twist angles of $1.538^\circ$, $2.004^\circ$ and $3.890^\circ$, the pressure-induced low-energy bands become exceptionally flat, with bandwidths below 10 meV. The bandwidths of these pressure-induced low-energy bands are comparable to that of the magic angle (1.084°) flatband. However, as the twist angle increases, achieving flatness in the low-energy bands becomes less feasible through pressure application. For the three higher twist angles, $6.008^\circ$, $7.340^\circ$ and $9.430^\circ$, the flattest bands achievable under perpendicular pressure are notably less flat than the flatband at the magic angle ($1.084^\circ$). Aside from band flatness, the four-band degeneracy at the Dirac point is lifted under pressure for higher twist angles as elaborated in Appendix~\ref{app:flat_low}. 

\begin{figure}[!ht!]
\begin{center}
    \phantomlabelabovecaption{(a)}{fig:bw_1}
    \phantomlabelabovecaption{(b)}{fig:bw_2}
    \includegraphics[width = 0.4\textwidth]{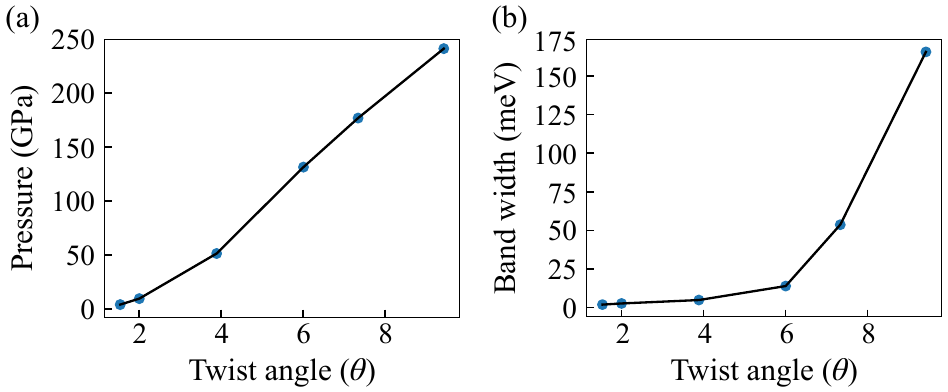}
    \caption{(color online)(a) Perpendicular pressure values in GPa (indicated by blue dots) required to reach the minimum bandwidth of the highest valence band for twist angles of $1.538^\circ$, $2.004^\circ$, $3.890^\circ$, $6.008^\circ$, $7.340^\circ$, and $9.430^\circ$, connected by a black line. (b) Bandwidth values of the highest valence band in meV (blue dots) at the point of minimum bandwidth under perpendicular pressure for the same twist angles, also connected by a black line.}
    \label{fig:bw}
\end{center}
\end{figure}
Further, we compare the bandwidth of the highest energy valence band at its minimum bandwidth point, along with the corresponding perpendicular pressure as a function of twist angle ($\theta$), in Fig.~\ref{fig:bw}. 
In Fig.~\ref{fig:bw_1}, we plot the pressure required to achieve the minimum bandwidth for each of these twist angles. As the twist angle increases, the perpendicular pressure required to reach the minimum bandwidth rises rapidly. This trend arises because larger twist angles weaken interlayer hybridization, requiring significantly higher pressures to restore strong coupling and induce band flattening. In Fig.~\ref{fig:bw_2}, we plot the bandwidth of the highest valence band at its minimum bandwidth under applied pressure. The bandwidth increases gradually up to $3.890^\circ$, remaining under 10 meV; however, it grows exponentially as the twist angle increases further. Thus, while perpendicular pressure can achieve a minimum bandwidth for higher twist angles, it does not produce the same level of flatness as observed at the magic angle. A natural question to be asked is, how these flat bands at each angle under pressure different from the magic angle flat band. In order to characterize them, we explored their spatially projected density of states (SPDOS), effect of magnetic field and obtained the Hall conductivity and confirm these topological phases from the Hofstadter Butterfly.   
\begin{figure*}[!ht!]
    \centering
    \phantomlabelabovecaption{(a)}{fig:spdos_1}
    \phantomlabelabovecaption{(f)}{fig:spdos_6}
    \phantomlabelabovecaption{(g)}{fig:spdos_7}
    \phantomlabelabovecaption{(h)}{fig:spdos_8}
    \includegraphics[width=0.9\linewidth]{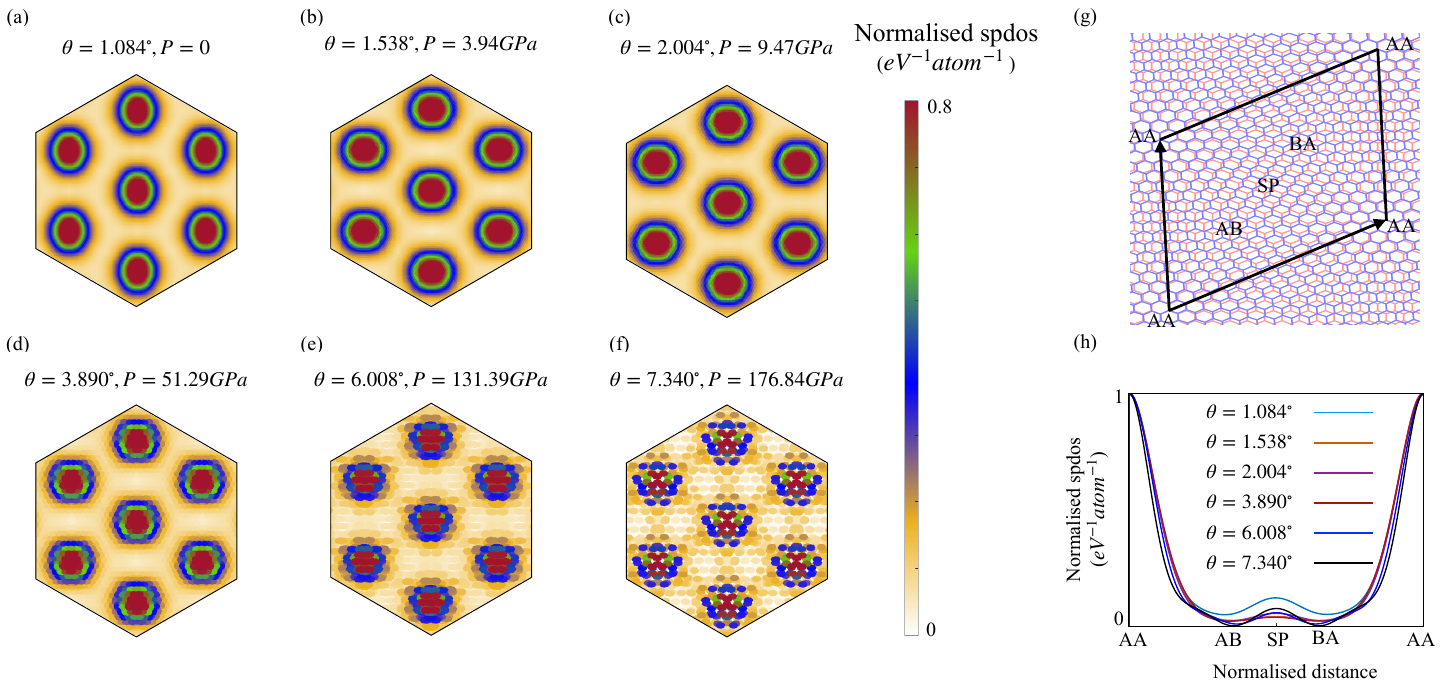}
    \caption{(color online)(a)-(f) Normalized spatial density of states (in units of $eV^{-1}atom^{-1}$) projected in real space at a specific energy level for the magic angle ($1.084^\circ$) without pressure and for higher twist angles under perpendicular pressure, where the bandwidth reaches its minimum.(g) Schematic of a moiré unit cell highlighting AA, AB, BA, and SP stacking regions along with their relative positions. (h) Normalized spatial density of states (in units of $eV^{-1}atom^{-1}$) along a path (shown in (g)) traversing all stacking regions in the moiré unit cell, shown for various twist angles at a specific energy level from the corresponding band structures in Appendix~\ref{app:flat_low}.}
    \label{fig:spdos}
\end{figure*}
\subsection{Spatial density of states under perpendicular pressure}\label{B}
The spatially projected density of states (SPDOS), describes the number of electronic states available at a specific energy level and localized at a specific spatial position in a material. Unlike the total density of states (DOS), which is averaged over the entire system, SPDOS provides information about how electronic states are distributed in real space. Mathematically, the spatial density of states at site $i$ is given by the diagonal elements of the imaginary part of the Green's function \cite{Souma}:

\begin{align}
    SPDOS(K, r_i,E) = -\frac{1}{\pi}Im G_{ii}(K,E)
\end{align}
where the retarded Greens function at energy E is:
\begin{align}
    G_{ij}(K,E) = \sum_{n}\frac{\psi_n^i(K) \psi_n^{j*}(K)}{E-\epsilon_n(K)+i\eta}
\end{align}
with $\eta$ is a small broadening parameter to avoid singularities, , $\epsilon_n(K)$ is eigenvalue(energy) of the n-th band and $\psi_m^i(K)$ is the component of eigenvector $\psi_n(K)$ at site $i$.\\
To obtain the SPDOS in real space, perform a discrete Fourier transform (sum over all K-points):
\begin{align}
    SPDOS(r_i,E) = \frac{1}{N_k}\sum_K SPDOS(K,r_i,E)
\end{align}
where $N_k$ is the number of $K$-points sampled in the Brillouin zone. To ensure comparability across different twist angles with varying unit cell sizes, SPDOS is normalized by the number of atoms in the unit cell:
\begin{align}
    SPDOS^{\textrm{Normalized}}(r_i,E) = \frac{SPDOS(r_i,E)}{\textrm{Number of atoms}}
\end{align}

In Fig.~\ref{fig:spdos_1}-\ref{fig:spdos_6} we present the normalized SPDOS projected in real space across the moiré unit cell for various twist angles at a specific energy, as indicated in Appendix~\ref{app:flat_low}. For the magic angle (1.084°), electronic states are highly localized in the AA regions, reflecting the strong localization observed in the SPDOS along the diagonal path of the unit cell. For higher twist angles, under the influence of perpendicular pressure at the point of minimum bandwidth, the electronic states exhibit a similar localization pattern, predominantly concentrated in the AA regions. This contrasts with the distribution of electronic states in the absence of perpendicular pressure, where they were spread more uniformly throughout the unit cell. The projection in real space provides a clearer visualization of this localization effect. Additionally, at lower twist angles, the SPDOS exhibits rotational symmetry around the AA regions, reflecting the moiré potential. However, at very high twist angles, this rotational symmetry is no longer preserved, indicating a reduced influence of moiré effects. 
\par The relative positions of the stacking orientations (AA, AB, SP, BA) in the twisted bilayer graphene unit cell are depicted in Fig.~\ref{fig:spdos_7}, with the origin set at an AA region. A diagonal path from one AA region to another traverses the AB, SP, and BA regions. The variation of normalized SPDOS is plotted along the diagonal path for different twist angles in Fig.~\ref{fig:spdos_8}. The distance between AA regions is normalized due to varying unit cell sizes, and the SPDOS height is normalized due to differing atom counts in the unit cells. At the magic angle, the AA region is maximally localized, with minimal (but non-zero) probability in the AB and BA regions, and a saddle point at SP. For higher twist angles (under perpendicular pressure), the AA region remains the most favorable for electronic states. However, the minimum SPDOS value decreases to zero at AB/BA regions as the twist angle increases, while a saddle point persists in the SP region. The SPDOS comparison of magic angle flat band and other pressured induced flat bands
indicates the similar localization scheme.
\subsection{Effect of perpendicular magnetic field on flatbands}\label{C}
\begin{figure*}[!ht!]
    \phantomlabelabovecaption{(a)}{fig:mag_1}
    \phantomlabelabovecaption{(b)}{fig:mag_2}
    \phantomlabelabovecaption{(c)}{fig:mag_3}
    \phantomlabelabovecaption{(d)}{fig:mag_4}
    \phantomlabelabovecaption{(e)}{fig:mag_5}
    \phantomlabelabovecaption{(f)}{fig:mag_6}
    \includegraphics[width = 0.9\textwidth]{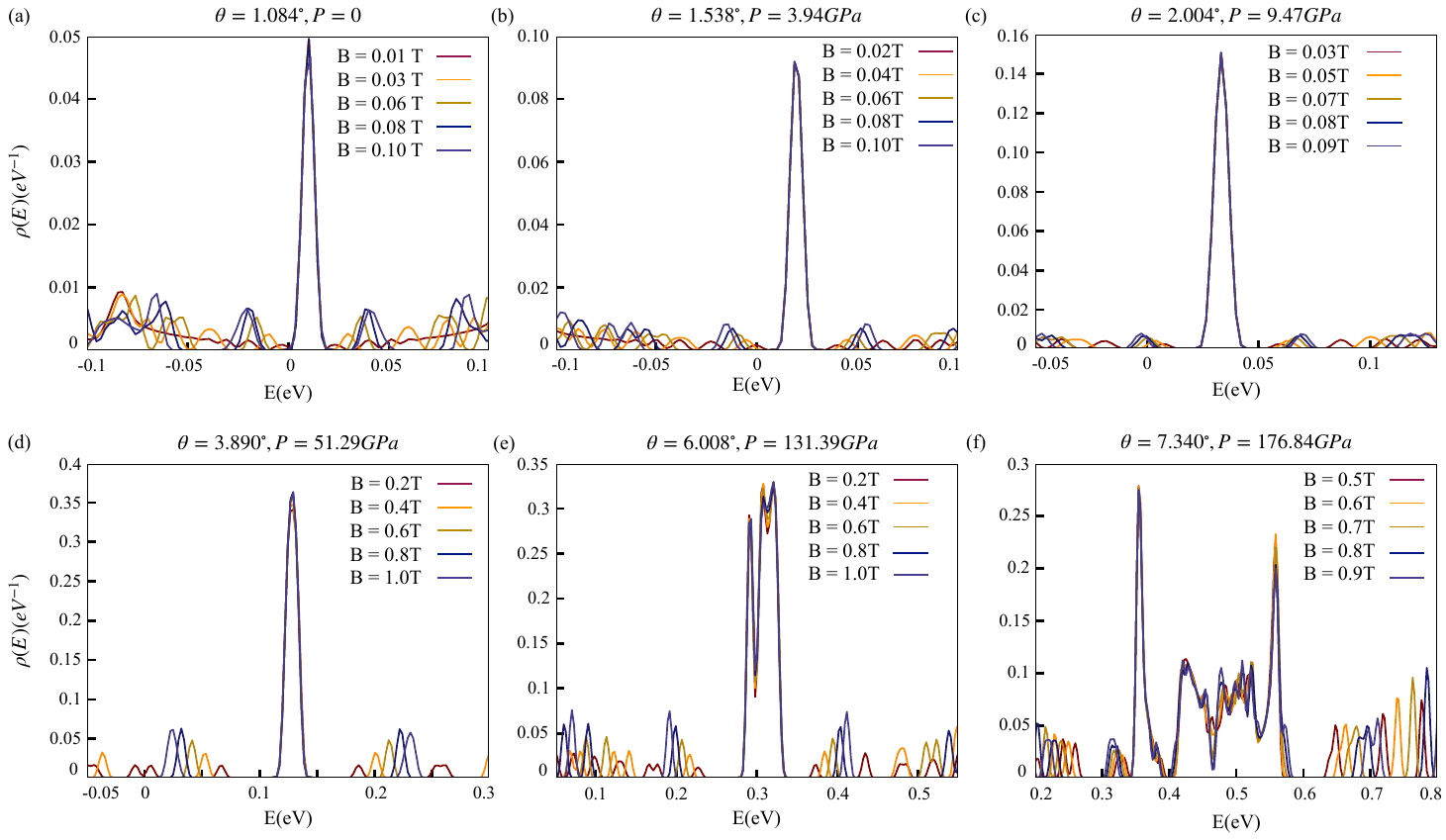}
    \caption{(color online)Density of states as a function of energy for various magnetic field strengths, ensuring the ratio $\phi / \phi_0 $ falls within the interval $35 \times 10^{-5}$ to $35 \times 10^{-4}$. The data is shown for different twist angles: $1.084^\circ$ (a), $1.538^\circ$ (b), $2.004^\circ$ (c), $3.890^\circ$ (d), $6.008^\circ$ (e), and $7.34^\circ$ (f), each under a uniform perpendicular pressure where the minimum bandwidth of the highest energy valence band is observed.}
    \label{fig:mag}
\end{figure*}
In this section, we discuss the effects of a magnetic field on the density of states for twisted bilayer graphene under pressure. For a homogeneous magnetic field $\Vec{B}=(0,0,B)$, where $B$ is the magnetic field strength, the hopping parameters become complex. Depending on the choice of the magnetic vector potential $\Vec{A}$ , a phase factor, known as the Peierls phase \cite{Peierls}, appears for hopping along certain paths. For an electrons hopping from site $i$ (position $\Vec{r}_i$) to site $j$ (position $\Vec{r}_j$) acquire the Peierls phase $exp(\frac{2\pi i}{\phi_0}\int_{\Vec{r}_i}^{\Vec{r}_j}\Vec{A}.d \Vec{l} )$ where $\phi_0 = \frac{hc}{e}$, where $\phi_0 = \frac{hc}{e}$ is the flux quantum, with $e$ as the electron’s charge, $h$ as Planck’s constant, and $c$ as the speed of light. Without loss of generality, we set $c=1$ throughout this paper. Under the application of a perpendicular magnetic field, the energy bands become discrete, forming what are known as Landau levels. For a finite lattice, the discreteness of these Landau levels depends on the ratio between the magnetic flux ($\phi = BS$) through the unit cell and the flux quantum ($\phi_0$), where $S$ is the area of the unit cell. In twisted bilayer graphene, the unit cell area ($S$) varies with the twist angle, meaning the magnetic field required to achieve well-separated Landau levels also depends on the twist angle.

\par In Fig.~\ref{fig:mag}, we plot the density of states for various twist angles under both perpendicular magnetic field and pressure. Since the moiré unit cell area changes with twist angle, we adjust the magnetic field strength to keep the flux ratio $\phi/\phi_0$ within a specified range for all twist angles. Previously~\cite{Sinha}, we have shown that for applied perpendicular magnetic field, Landau levels (LLs) generated from the remote bands but does not alter the flatband (Fig.~\ref{fig:mag_1} at the magic angle. For $B=0.01T$ ($\frac{\phi}{\phi_0} = 35\times 10^{-5}$), the LLs are not well-separated, but as the magnetic field strength increases, the separation between LLs becomes more pronounced. At $B=0.1T$ ($\frac{\phi}{\phi_0} = 35\times 10^{-4}$), the LLs are clearly separated up to the first order, although higher-order LLs are influenced by bands from the edges. However, across this entire range of $B$, the density of states for the flatbands remains unchanged. This stability is due to the topological protection of the flatbands, as the magnetic field $B$ cannot break the $C_3$ symmetry inherent in the system \cite{Sinha, Gail}.
\begin{figure*}[!ht!]
    \phantomlabelabovecaption{(a)}{fig:hall_1}
    \phantomlabelabovecaption{(b)}{fig:hall_2}
    \phantomlabelabovecaption{(c)}{fig:hall_3}
    \phantomlabelabovecaption{(d)}{fig:hall_4}
    \phantomlabelabovecaption{(e)}{fig:hall_5}
    \phantomlabelabovecaption{(f)}{fig:hall_6}
    \includegraphics[width = 0.9\textwidth]{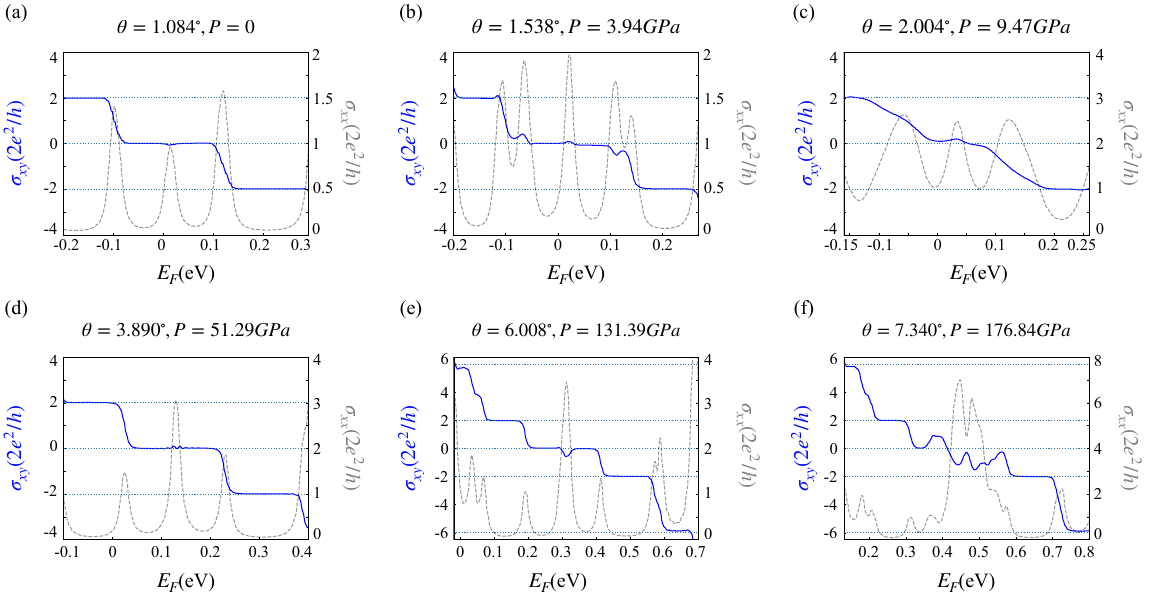}
    \caption{(color online)Hall conductivity ($\sigma_{xy}$) and longitudinal conductivity ($\sigma_{xx}$) (in units of $2e^2/h$) is plotted as a function of Fermi energy, $E_F$ (in units of eV) for different twist angles $1.084^{\circ}$ (a), $1.538^{\circ}$ (b), 
    $2.004^{\circ}$ (c),
    $3.890^{\circ}$ (d), $6.008^{\circ}$ (e), and $7.340^{\circ}$ (f) for a fixed uniform magnetic field ($B=1$ T) and uniform perpendicular pressure as mentioned.}
    \label{fig:hall}
\end{figure*}
\par We now examine the effects of a perpendicular magnetic field on higher twist angles at the point of minimum bandwidth for the lowest energy bands, achieved under perpendicular pressure. For all cases, we ensure that the ratio $\phi/\phi_0$ falls within the same range as for the magic angle. We observe that a perpendicular magnetic field can generate well-resolved first-order Landau levels (LLs) without altering the density of states for the lowest energy bands. This holds for twist angles of $1.538^\circ$ (Fig.~\ref{fig:mag_2}), $2.004^\circ$ (Fig.~\ref{fig:mag_3}), and $3.890^\circ$ (Fig.~\ref{fig:mag_4}). However, as we move to higher twist angles, the bandwidth of the lowest energy bands at their minimum increases, causing the van Hove singularity to diminish as these bands become more dispersive. Even at higher twist angles, unaffected low-energy bands remain, as the twist does not disrupt $C_3$ symmetry. At $6.008^\circ$ (Fig.~\ref{fig:mag_5}), the flatness in the low-energy bands is lost, but they remain unaffected by the perpendicular magnetic field, with visible first-order LLs on both sides. At $7.340^\circ$ (Fig.~\ref{fig:mag_6}), the width of the low-energy bands becomes substantial, ranging from $0.35$ eV to $0.57$ eV. A closer examination of this region reveals that while some peaks within the four low-energy bands shift under the magnetic field, the boundaries remain stable. Due to particle-hole symmetry breaking (as shown in the band structure), the first-order LLs are asymmetric on both sides, with LL separations differing between the positive and negative energy regions. In the case of pressure induced flat bands, the flat band is unaltered by the magnetic fields and the driven LLs are mainly from the remote bands. 

\subsection{Effect of pressure on Hall Conductivity}\label{D}
We compute the DC conductivity tensor in the linear-response regime using the real-space Kubo-Bastin \cite{Kubo1,Kubo2} formalism, as implemented in the KITE \cite{João,Pires1,Pires2} code. This approach enables large-scale quantum transport calculations for TBG systems, capturing moiré-scale physics while maintaining numerical efficiency. In Fig.~\ref{fig:hall}, we present the Hall conductivity and longitudinal conductivity, both expressed in units of $2e^2/h$ (accounting for spin degeneracy), under a fixed uniform perpendicular magnetic field of $B=1T$ for various twist angles. Fig.~\ref{fig:hall_1} shows the results for the magic angle ($1.084^\circ$) with zero pressure, while Fig.~\ref{fig:hall_2}-\ref{fig:hall_6} correspond to higher twist angles ($1.538^\circ, 2.004^\circ, 3.890^\circ, 6.008^\circ $ and $7.340^\circ$) under a uniform perpendicular pressure, chosen such that the lowest energy bandwidth reaches its minimum. 

\par For the magic angle ($1.084^\circ$), the Hall conductivity, $\sigma_{xy}$, exhibits well-defined plateaus at 0 and $\pm~2$ in units of $2e^2/h$ under a perpendicular magnetic field of $B=1T$. The longitudinal conductivity($\sigma_{xx}$) remains nonzero at the transition points between these plateaus, where electronic states contribute to dissipative transport. However, in regions where the Hall conductivity forms a zero plateau, $\sigma_{xx}$ develops a distinct peak, indicating an increased density of states and enhanced scattering processes. For higher twist angles, the Hall conductivity follows the same pattern as that of the magic angle, a behavior that was absent without the application of perpendicular pressure. For twist angles $1.538^\circ$, $2.004^\circ$ and $3.890^\circ$ under uniform pressure, the Hall conductivity plateaus appear at 0, $\pm~ 2$ , albeit with some fluctuations. The corresponding longitudinal conductivity becomes nonzero at the transition points between plateaus and within the regions of zero Hall conductivity. At even higher twist angles, specifically $6.008^\circ$ and $7.340^\circ$ , additional higher-order plateaus emerge alongside the 0 ,$\pm~ 2$ plateaus for the same magnetic field ($B=1T$). Due to the absence of electron-hole symmetry at very high twist angles under perpendicular pressure, these higher-order Hall plateaus are not symmetric on both the electron and hole sides. For a twist angle of $6.008^\circ$, the Hall plateaus at $+4$, $+6$ and $-6$ become noticeable and could become even more distinct with an increase in the magnetic field strength. Similarly, for a twist angle of $7.340^\circ$ , the Hall plateaus at $+6$, $+4$, $+2$, $-2$ and $-6$ are distinguishable, with some fluctuations appear near the zero Hall plateau. Importantly, we conclude that the Hall conductivity of the pressure induced flat bands have demonstrated  an integer quantum Hall effect $\sigma_{xy} = \pm~ 2n(2e^2/h)$ ($n=0,\pm~ 1,\pm~ 2,…$) same as magic agle flat band with a well-defined zero-energy Hall plateau appears.
\begin{figure*}[!ht!]
    \phantomlabelabovecaption{(a)}{fig:fly_1}
    \phantomlabelabovecaption{(b)}{fig:fly_2}
    \phantomlabelabovecaption{(f)}{fig:fly_6}
    \includegraphics[width = 0.9\textwidth]{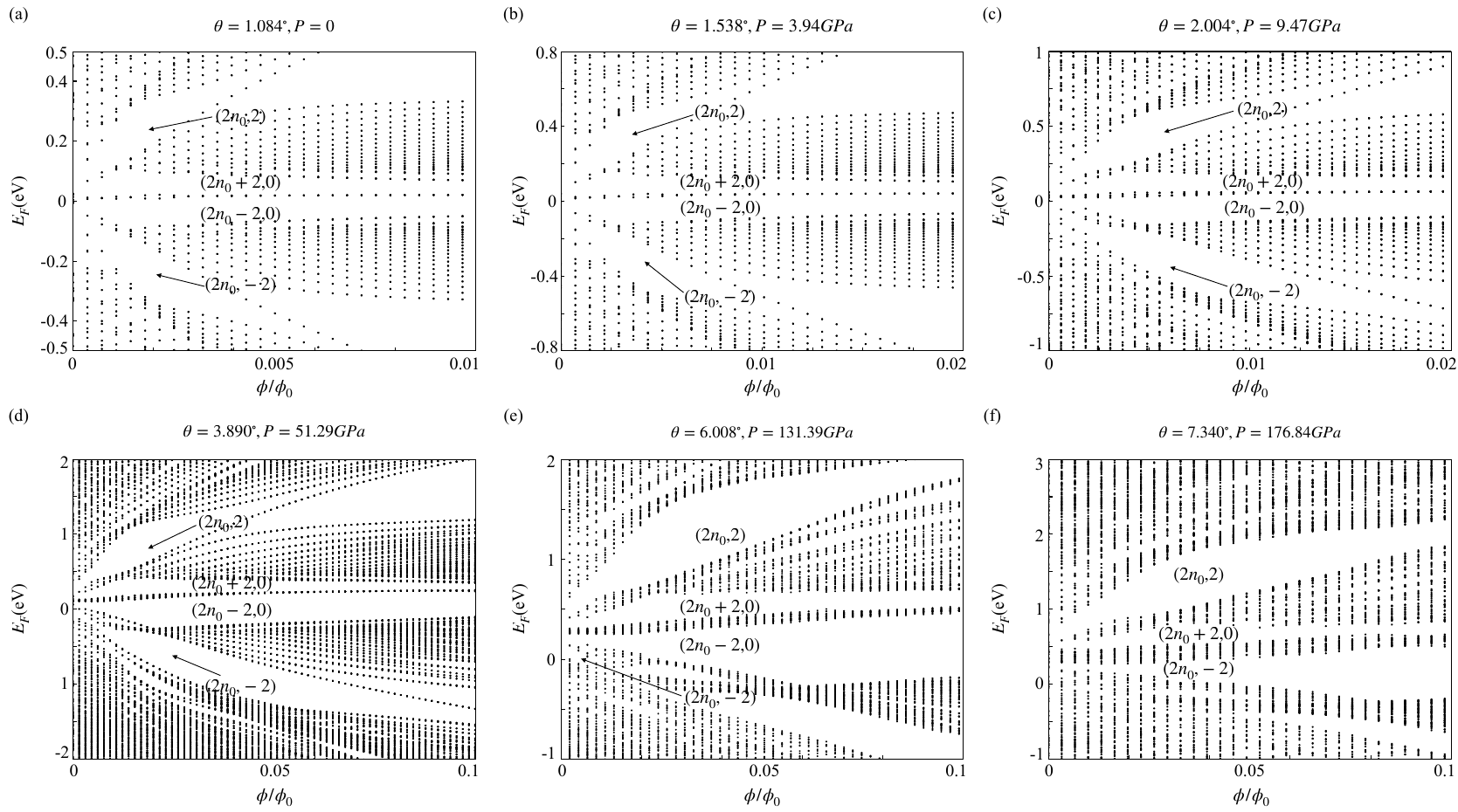}
    \caption{Close view of energy spectrum ($E_F(eV)$) as a function of flux per unit cell ($\phi/\phi_0$) for magic angle ($1.084^\circ$) (a) and higher twist angles $1.538^\circ$ (b), $2.004^\circ$ (c), $3.890^\circ$ (d), $6.008^\circ$ (e), $7.340^\circ$ (f) under perpendicular pressure.}
    \label{fig:fly}
\end{figure*}
\subsection{Hofstadter Butterfly under perpendicular pressure}\label{E}
Further, we present the Hofstadter butterfly diagram (energy versus magnetic field) for twisted bilayer graphene under uniform perpendicular pressure. As explained in this reference \cite{Hasegawa}, we adopt the periodic Landau gauge instead of the conventional Landau gauge. The periodic Landau gauge maintains periodicity in one direction, making it possible to obtain the Hofstadter butterfly diagram. For the applied magnetic field in a system, the flux per supercell is given by $\Phi = \frac{p}{q} \phi_0$, where $p$ and $q$ are mutually prime integers and $\phi_0$ is the magnetic flux quantum. The available energy levels are obtained as the eigenvalues of a $4n_0q \times 4n_0q$ matrix. This results in $4n_0q$ energy levels separated by $4n_0q-1$ gaps. For the chemical potential lies in the $r$th gap from the bottom, the system follows the Diophantine equation :
\begin{equation}
    r=qa_r+pb_r
\end{equation}
where $a_r$ and $b_r$ are integer coefficients. The Hall conductance is directly related to $b_r$ and is given by \cite{Thouless,Kohmoto1, Kohmoto2}:
\begin{equation}
    \sigma = 2 \frac{e^2}{h} b_r
\end{equation}
 where the factor of 2 accounts for spin degeneracy. When the Fermi energy lies within a gap, the Hall conductivity ($\sigma_{xy}$) is quantized, forming a plateau. In this state, the conduction occurs only through edge states. As the Fermi energy shifts from one gap to another, the system undergoes a transition. During this transition, the Fermi energy moves through a region of extended states within the energy bands, leading to a peak in the longitudinal conductivity.
 \par We have presented the full range of Hofstadter Butterfly obtained for all the twist angles for the pressures when the flat bands are induced in the Appendix:~\ref{app:full_fly}. In Fig.~\ref{fig:fly}, we plot the Fermi energy ($E_F$) as a function of the magnetic flux per unit cell in each layer ($\phi = \frac{\Phi}{n_0}$) for the magic angle Fig.~\ref{fig:fly_1} and higher twist angles (Fig.~\ref{fig:fly_2}-\ref{fig:fly_6}) under perpendicular pressure. For a continuous gap, the values of ($a_r,b_r$) remain constant. Here, we focus on the gaps that persist down to very low magnetic fields. For the magic angle ($1.084^\circ$), we observe four gaps in the low-field region characterized by $b_r = -2,0,2$, which is consistent with the Hall conductivity obtained in the previous section.  Notably, there are two distinct gaps with $b_r=0 $, each corresponding to different values of $a_r$, separated by intermediate energy bands. These bands act as extended states, resulting in a nonzero longitudinal conductivity within the zero Hall plateau. For higher twist angles ($1.538^\circ$,$2.004^\circ$, $3.890^\circ$ and $6.008^\circ$) under perpendicular pressure, we also observe four distinct energy gaps characterized by $b_r=-2,0,2$. Similar to the magic angle case, there are extended states present between the two $b_r=0$ gaps, which contribute to nonzero longitudinal conductivity in the corresponding regions. For twist angle $7.340^\circ$, the identified gaps correspond to $b_r=-2,0,2$ with a single gap for $b_r=0$. Towards very high twist angles ($6.008^\circ$ and $7.340^\circ$), broken electron-hole symmetry becomes more pronounced. 
 \begin{figure}[!ht!]
\begin{center}
    \phantomlabelabovecaption{(a)}{fig:flat_band1}
    \phantomlabelabovecaption{(b)}{fig:flat_band2}
    \phantomlabelabovecaption{(c)}{fig:flat_band3}
    \includegraphics[width = 0.4\textwidth]{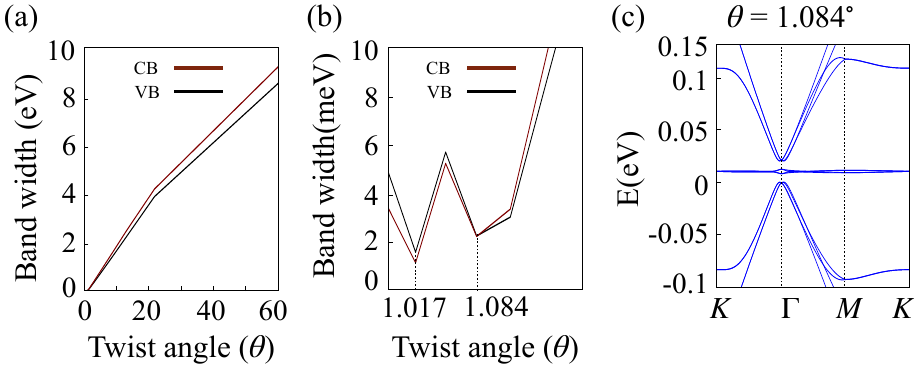}
    \caption{(color online)(a) Variation in the bandwidth of the highest energy valence band (black) and the lowest energy conduction band (brown) as a function of the relative twist angle between two graphene layers. (b) A closer view of the low twist angle region ($<3^\circ$) from the previous plot, highlighting the highest minimum in bandwidth at $1.084^\circ$ , which is the largest angle at which low-energy flatbands appear. (c) Band structure of twisted bilayer graphene in the moiré Brillouin zone for a twist angle of $1.084^\circ$, shown along the high-symmetry path $K-\Gamma-M-K$. }
    \label{fig:flat_band}
\end{center}
\end{figure}
\section{Conclusions}\label{IV}
In summary, we investigated the effect of perpendicular pressure on large-angle twisted bilayer graphene (TBG) and its ability to induce flatbands, revealing important modifications in electronic structure, localization, transport, and quantum Hall behavior. We found that pressure effectively reduces the bandwidth at higher twist angles, though the degree of flattening diminishes as the angle increases, with bands becoming more dispersive beyond $\sim 6^\circ$. At larger angles, the four-band degeneracy at the Dirac point is broken, reducing it to a two-band degeneracy, indicating a transition toward a decoupled bilayer regime. The spatially projected density of states (SPDOS) revealed strong electron localization in the AA-stacked regions under pressure, whereas without pressure, states remained more uniformly distributed. Interestingly, SPDOS exhibited rotational symmetry around the AA regions at intermediate angles, but this symmetry was lost at very high angles, signifying a weakening of moiré effects. The application of a perpendicular magnetic field showed that pressure-induced flatbands at lower twist angles behaved similarly to the magic-angle case, where Landau levels (LLs) emerged without significantly affecting the density of states. However, at very high angles, increasing bandwidth and the breaking of electron-hole symmetry led to asymmetric LL formation. Hall conductivity calculations showed that pressure-induced flatbands at moderate angles displayed similar quantized plateaus as in magic-angle TBG, but at very high angles, additional plateaus appeared, and electron-hole symmetry was lost. The Hofstadter spectrum shows four consistent low-energy gaps across all twist angles under pressure, where $b_r$ denotes the Hall plateaus. These gaps align with Hall conductivity plateaus, while the two $b_r=0$ gaps explain the kinks in the zero Hall plateau. These results demonstrate that while perpendicular pressure can extend magic-angle-like behavior to larger twist angles, there is a fundamental limit beyond which moiré-induced electronic correlations weaken. Our findings highlight pressure as a powerful tool for tuning electronic properties in TBG and pave the way for further studies on pressure-engineered moiré systems and their quantum transport applications.
\begin{figure}[!ht!]
\begin{center}
    \phantomlabelabovecaption{(a)}{fig:low_energy1}
    \phantomlabelabovecaption{(f)}{fig:low_energy2}
    \includegraphics[width = 0.4\textwidth]{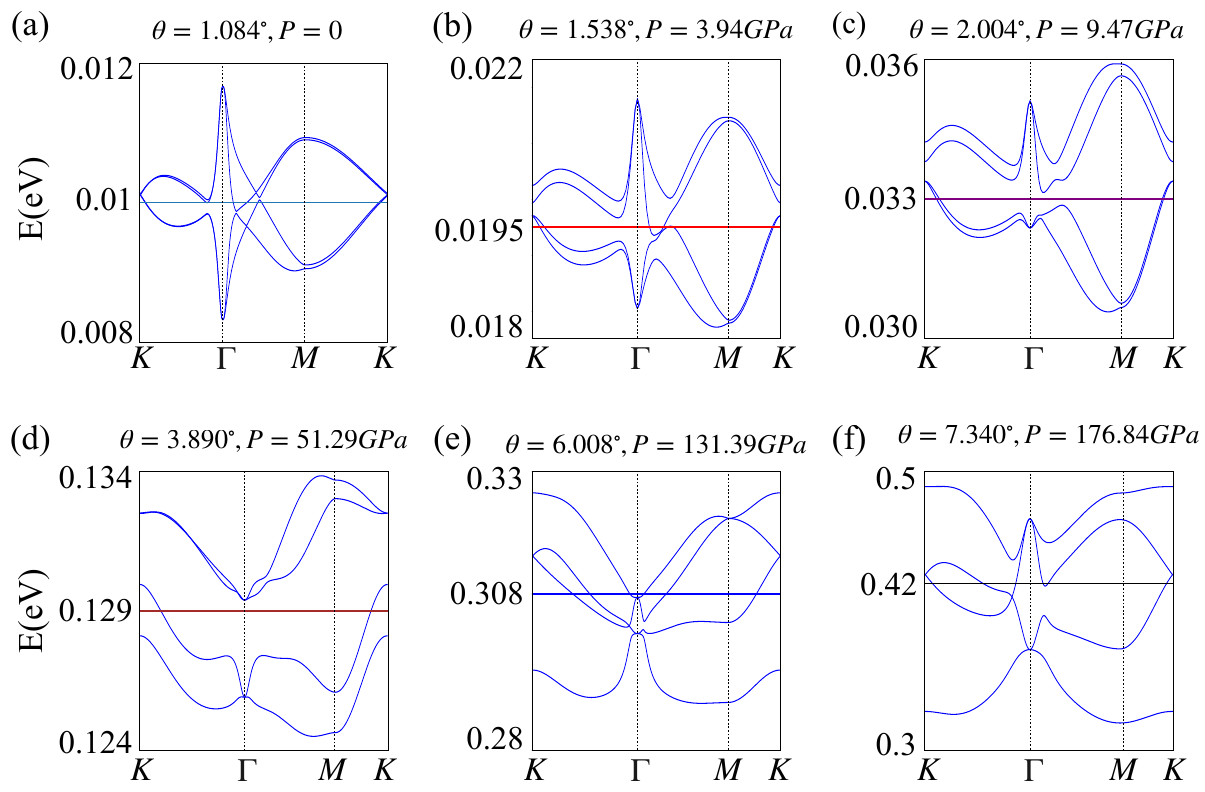}
    \caption{(color online)(a)-(f) Lowest energy bands for the magic angle ($1.084^\circ$) without pressure and for higher twist angles under perpendicular pressure, where the bandwidth reaches its minimum.}
    \label{fig:low_energy}
\end{center}
\end{figure}
\begin{figure*}[!ht!]
\begin{center}
    \includegraphics[width = 0.9\textwidth]{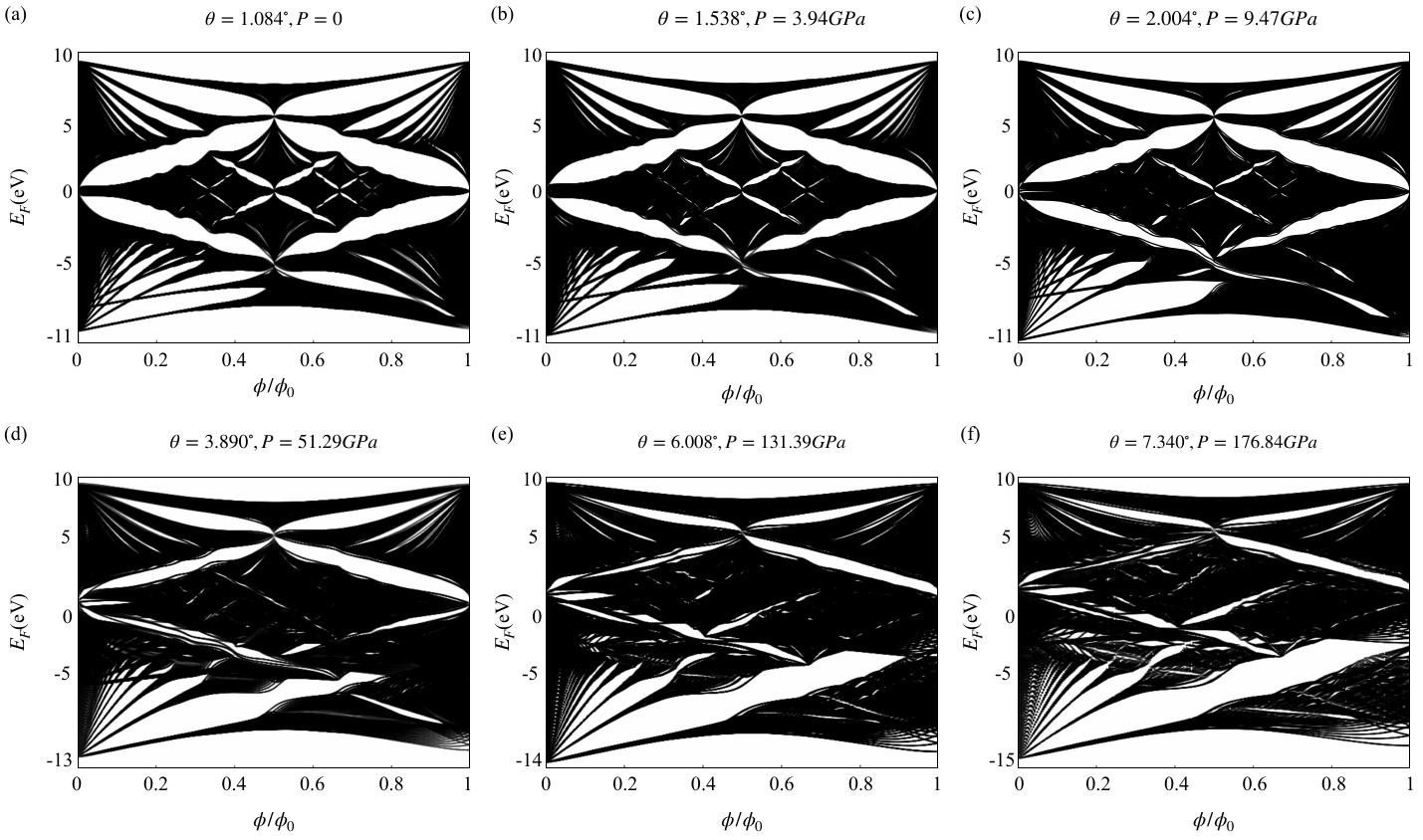}
    \caption{Energy spectrum($E_F$(eV)) of twisted bilayer graphene as a function of magnetic flux per unit cell ($\phi / \phi_0$) for different twist angles under uniform perpendicular pressure.}
    \label{fig:butterfly}
\end{center}
\end{figure*}
\section*{ACKNOWLEDGMENTS}
The authors acknowledge the support provided by the KEPLER computing facility, maintained by the Department of Physical Sciences, IISER Kolkata. A.M. and P.S. acknowledge the financial support from IISER Kolkata through the Institute PhD and Post-doctoral Fellowship, respectively. B.L.C. acknowledges the SERB for Grant No. SRG/2022/001102 and “IISER Kolkata Start-up-Grant” Ref. No. IISER-K/DoRD/SUG/BC/2021-22/376.
\appendix
\section{ Magic angle flatband}\label{app:magic_flat}

In our model, we begin by placing two layers of graphene directly on top of each other in an AA stacking configuration before twisting the top layer. It is well known that as the twist angle decreases, the area of the moiré unit cell increases, while the width of the lowest energy bands (specifically, the highest valence band and the lowest conduction band) decreases. At the commensurate twist angle near $1.1^\circ$—known as the "magic angle"—these low-energy bands become extremely flat, on the order of a few meV. Our model successfully demonstrates the decrease in bandwidth with decreasing twist angle and reveals the flat band that emerges at the commensurate angle of $1.084^\circ$, as shown in Fig.~\ref{fig:flat_band}. In Fig.~\ref{fig:flat_band1}, we plot the bandwidths of the two lowest energy bands, with black representing the highest valence band and brown the lowest conduction band. The bandwidth declines sharply as the twist angle decreases, with flat bands appearing when the bandwidth reaches a minimum. The first occurrence of a flat band at the highest minima in bandwidth aligns with the commensurate angle of $1.084^\circ$, shown in Fig.~\ref{fig:flat_band2}. In Fig.~\ref{fig:flat_band3}, we plot the band structure in momentum space for the twist angle of $1.084^\circ$. Here, four low-energy bands are observed due to contributions from two sublattices and two layers, assuming a single-spin system. These four bands meet at the corners of the superlattice Brillouin zone, specifically at the Dirac points. Additionally, the low-energy bands are well-separated from the higher-energy bands on both the electron and hole sides.

\section{Pressure induced low energy bands}\label{app:flat_low}

In Fig.~\ref{fig:low_energy1}-\ref{fig:low_energy2} the four lowest energy bands are shown for the magic angle (1.084°) and pressure-induced higher twist angles ($1.538^\circ$, $2.004^\circ$, $3.890^\circ$, $6.008^\circ$ and $7.340^\circ$), where the bandwidth reaches a minimum.  As the twist angle increases, particle-hole symmetry gradually breaks in the low-energy bands. At certain intermediate angles, a splitting of the low-energy bands is observed with the presence of an indirect bandgap. Though in all cases, the bands are not splitting apart at the $\Gamma$ point. This preserves the $C_3$ symmetry. But, the band touching at the Dirac point is different in all higher twist angles. In contrast to the magic angle and the pressure-induced lower twist angles, not all low  energy bands touch each other at Dirac point reflecting the broken inversion symmetry ($C_2$).  A straight line within the band structure plot highlights the specific energy near the middle of the lowest energy region, chosen for SPDOS calculations.

\section{Pressure induced Hofstadter Butterfly}\label{app:full_fly}
In Fig.~\ref{fig:butterfly}, we present the Hofstadter butterfly spectrum (energy spectrum ($E_F(eV)$) as a function of magnetic field per unit cell ($\phi/ \phi_0$)) for the magic angle ($1.084^\circ$) and higher twist angles ($1.538^\circ$, $2.004^\circ$, $3.890^\circ$, $6.008^\circ$, and $7.340^\circ$) under perpendicular pressure. Each gap in the spectrum is characterized by two integers ($a_r$, $b_r$) corresponds to the Hall quantization plateaus. The application of pressure modifies these gaps, leading to the emergence of new quantization plateaus. In this study, we focus on the low-energy and low-magnetic-field regions, as highlighted in Fig.~\ref{fig:fly}, where pressure-induced quantization levels are evident. Notably, these newly formed plateaus align well with the Hall conductivity plateaus shown in Fig.~\ref{fig:hall}.

\end{document}